\pdfoutput=1
\documentclass{jdfsl}

\usepackage{makeidx}  
\usepackage{upgreek}

\usepackage{url}

\usepackage{amsmath}

\usepackage[T1]{fontenc}

\usepackage{array}

\usepackage{subfig}
\usepackage{multirow}
\usepackage{array}
\usepackage{amssymb}

\usepackage{natbib}

\usepackage{enumitem}

 \usepackage{amssymb}
 \usepackage{pifont}
\usepackage{textcomp}

\usepackage{upgreek}

\usepackage{url}

\usepackage{amsmath}
\usepackage{graphicx}

\usepackage[T1]{fontenc}

\usepackage{array}
\graphicspath{{./pdf/}{./jpeg/}{./images/}}
   \DeclareGraphicsExtensions{.pdf,.jpeg,.jpg,.png}
   
\usepackage{eurosym}
\begin{document}

\title{Project Maelstrom: Forensic Analysis of the BitTorrent-Powered Browser}


\author{Jason Farina, M-Tahar Kechadi, Mark Scanlon}
\institute{School of Computer Science, University College Dublin, Ireland. \\ jason.farina@ucdconnect.ie, \{tahar.kechadi, mark.scanlon\}@ucd.ie}

\abstract{
In April 2015, BitTorrent Inc. released their distributed peer-to-peer powered browser, Project Maelstrom, into public beta. The browser facilitates a new alternative website distribution paradigm to the traditional HTTP-based, client-server model. This decentralised web is powered by each of the visitors accessing each Maelstrom hosted website. Each user shares their copy of the website;s source code and multimedia content with new visitors. As a result, a Maelstrom hosted website cannot be taken offline by law enforcement or any other parties. Due to this open distribution model, a number of interesting censorship, security and privacy considerations are raised. This paper explores the application, its protocol, sharing Maelstrom content and its new visitor powered ``web-hosting'' paradigm.
}

\maketitle

\section{Introduction}
\label{intro}

Project Maelstrom was released as a private alpha in December 2014 \citep{klinker2014project} and as a public beta in April 2015 \citep{klinker2015project}. Its purpose is to provide a decentralised web ecosystem facilitating a new parallel to the existing world wide web. Through this decentralisation, users are free to create and share any content they desire without the need for web hosting providers or domain names, and can bypass any national or international censorship. The HTML documents, associated styling and scripting files, multimedia content, etc., are hosted by the website's visitors, and subsequently served to other visitors accessing the site. The official BitTorrent blog celebrated its arrival as: 
\begin{quote}
\vspace{-0.2cm}\textit{Truly an Internet powered by people, one that lowers barriers and denies gatekeepers their grip on our future} \citep{klinker2014project}. 
\end{quote}
\vspace{-0.2cm}
While the topic of BitTorrent in the media seems to predominantly coincide with a discussion on online piracy \citep{choi2007online}, the protocol has proven itself as a robust, low-cost, distributed alternative to the traditional client-server content distribution model. From a forensic standpoint, the decentralised nature of BitTorrent based applications can result in extended windows for evidence acquisition \citep{scanlon2014leveraging}. Aside from facilitating copyright infringement, a number of additional online services have been developed using the protocol including:

\begin{itemize}[noitemsep,topsep=0pt]
\item File Synchronisation Service -- BitTorrent Sync is a cloudless alternative to the cloud-based, multiple-device, file synchronisation services such as Dropbox, OneDrive, iCloud, etc. \citep{Scanlon2015}. With a standard BitTorrent Sync install, users are not limited in the amount of data they can share as there are no replicated server-side limitations.
\item Cost-Effective Commercial Content Distribution -- BitTorrent Inc. use the protocol to distribute commercial multimedia content through their ``BitTorrent Bundle'' offering. Large video game creators and distribution companies, Blizzard Entertainment and Valve have used the BitTorrent protocol to distribute installation files and software patches to their users \citep{Watters201179}. The advantage for these companies is that the protocol excels where the traditional client-server model starts to fail; the more users, the faster the average download speed.
\end{itemize}

\subsection{Contribution of this Work}
\label{contribution}
The contribution of this work can be summarised as:

\begin{itemize}[noitemsep,topsep=0pt]
\item An overview and analysis of Project Maelstrom's functionality.
\item An analysis of the forensic value of its installation and configuration files.
\item An investigation of evidence left behind after uninstallation and what knowledge be recovered.
\end{itemize}

\section{Background Reading}
\label{background}

\subsection{Peer-to-Peer Facilitated Web Browsing}
The deep web refers to layers of Internet services and communication that is not readily accessible to most users and is not crawlable by traditional search engines. Unlike the regular Internet, there is not one set of protocols or formats for the deep web, instead deep web is a generic term used to describe Internet communications that are managed using closed or somehow restricted protocols. More recently however, the term ``deep web'' has been made synonymous with black-market sites such as ``The Silk Road'', taken offline by the FBI in 2013. While BitTorrent Inc. do not claim that Project Maelstrom enhances privacy or anonymises a user's Internet traffic, many of the competing decentralised deep web technologies focus their efforts on precisely that.

\subsubsection{The Freenet Project}

The Freenet project is a peer based distributed Internet alternative. Users connect to Freenet through an installed application that uses multiple encrypted connections to mask the identifying network information and the data location. Freenet peers, or nodes, store fragments of data in a distributed fashion. The number of times a data item is replicated is dependent on the demand for that data. More popular files have more available sources resulting in better availability and faster access times for the requester. The layered encryption of the connections provides an effective defence against network sniffing attacks and also complicates network forensic analysis. Freenet, by default, is deployed as an OpenNet service. On OpenNet, it is possible to enumerate and connect to all available nodes and perform traffic correlation to trace back to the original requester \citep{roos2014measuring}.

\subsubsection{Tor}

Tor (The Onion Router) is a networking protocol designed to provide anonymity while accessing both regular websites, and those hosted within the Tor network. Initially, Tor provided random internal encrypted routes enabling anonymised regular website access for its users. This was achieved through a minimum of three encrypted connections before exiting to the regular Internet through an exit node. 

A user can opt not to utilise an exit node and instead browse content on services hosted within the Tor network itself. These Tor websites are known as Hidden Services (HS) and the identity and geolocation of the hosting server are obfuscated, in a similar fashion to the website's visitors. This anonymisation would normally make any form of dependable routing impossible as any DNS lookup would not be able to report a static IP address for the service for the client system to connect to. Instead, hidden services use a distributed hash table (DHT) and a known service advertising node to inform clients of their presence and the path to the service provided.

Due to the reliance of the HS on DHT and directory servers to perform the introductions for Tor users, \cite{Biryukov2013} were able to demonstrate a denial of service attack on a HS by impersonating the directory servers. They were also able to crawl the DHT to harvest Onion Identifiers over a period of two days, resulting in an accurate index of the content of the deep web contained within Tor. This approach could be useful when attempting to stall a HS until it can be properly identified.

\subsubsection{I2P}
The Invisible Internet Project (I2P) is often considered another anonymisation utility like TOR. However, its garlic routing protocol is intended to create an alternative Internet. In this model, all traffic takes place via unidirectional tunnels established out to a resource which responds with its own, different, tunnelled return path. In this way all traffic to and from a host is protected by a separate session based tunnel that closely resembles that of TOR, but I2P was not intended to allow access to sites or services outside of the I2P network itself. In this regard, I2P more closely resembles Tor Hidden Services. In I2P, the traffic is encrypted end-to-end and the encryption in either direction is handled separately \citep{zantout2011i2p}.

\section{Project Maelstrom: Application Analysis}
\label{application}

\begin{figure}[!t]
\centering
\includegraphics[width=0.5\textwidth]{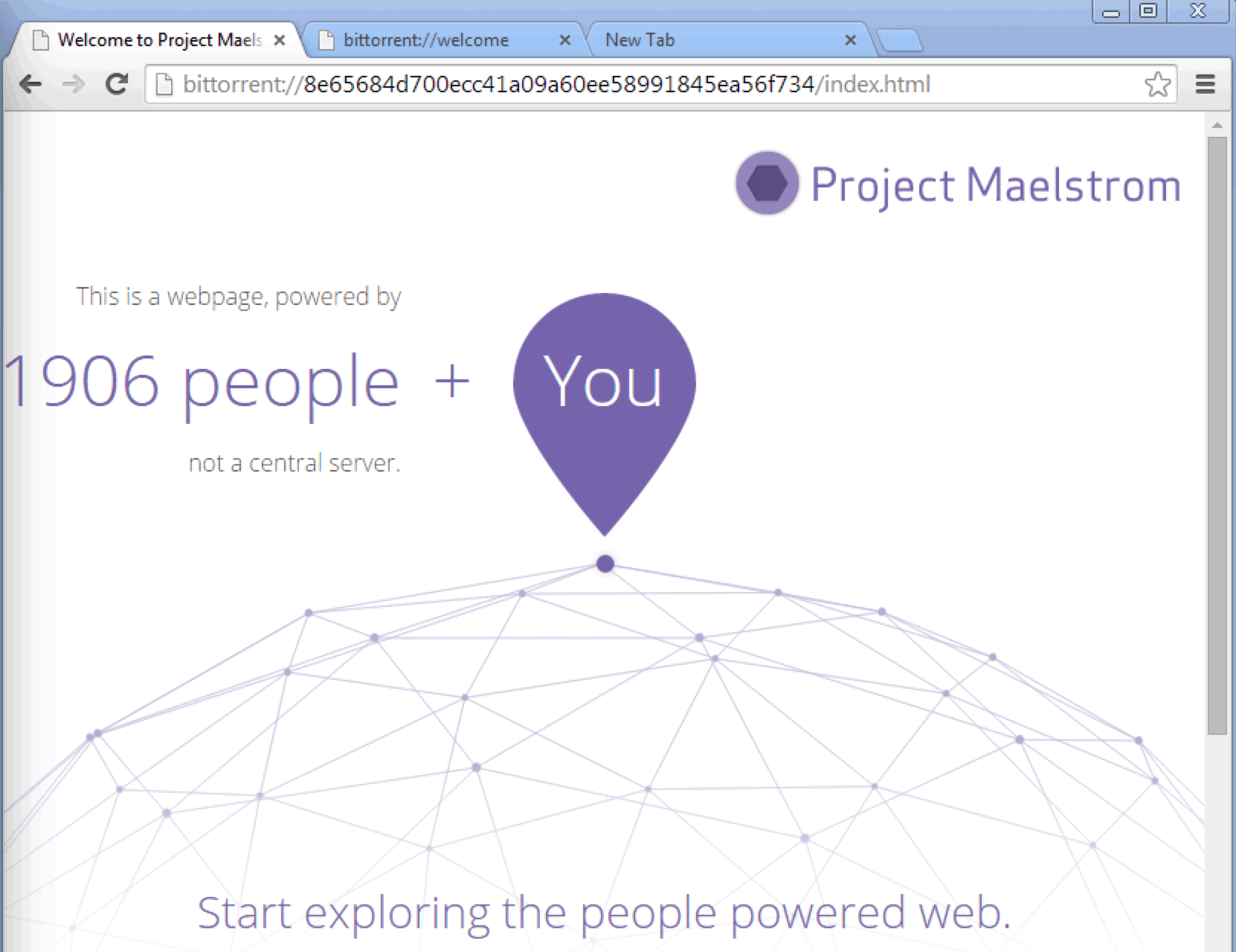}
\caption{Project Maelstrom Default Start Page}
\label{fig:startpage}
\end{figure}

Project Maelstrom is built upon the open source version of Google's Chrome browser, Chromium, as can be seen in Figure~\ref{fig:startpage}.

\subsection{Installation}
\label{installation}
The version of Project Maelstrom analysed as part of this paper has the following properties\\
\texttt{File:} Maelstrom.exe\\
\texttt{Size:} 36,971KB\\
\texttt{MD5:} d3b6560c997a37a1359721fdaa25925f\\
\texttt{SHA1:} 9de7c48e324b2ac9240ca168b7c2afd46bd1c799\\
\texttt{Origin:} download-lb.utorrent.com\\
\texttt{version:} 37.0.2.1

All testing was carried out on a virtual machine running Windows 7 with 1GB ram and a thin provisioned 60gb hard drive.

The Project Maelstrom installation process is in two parts. The first part installs a browser based on Chrome that acts as a user interface for the Maelstrom network. The second is the BitTorrent Maelstrom application that handles caching of torrents and maintains connection to the distributed hash table (dht). 

Once executed, the Project Maelstrom installer extracts the required installation files to  a \texttt{TEMP} directory in \texttt{\textbackslash Users \textbackslash <username> \textbackslash Appdata}. Chrome.7z and setup.exe are stored in a directory created at \texttt{ \textbackslash Local\textbackslash Maelstrom\textbackslash Application\textbackslash <version number>\textbackslash Installer}. 

All pre-defined URLs used during installation are stored within settings files extracted from the Chrome.7z package and IP addresses are resolved as required through standard DNS queries. This allows the source servers to change addresses without invalidating older installation packages. The Project Maelstrom specific URLs at the time of writing are: \url{https://s3-us-west-1.amazonaws.com}, \\\url{router.bittorrent.com}, \\\url{update.browser.bittorrent.com}, \\\url{router.utorrent.com}, \\\url{bench.utorrent.com}, \\\url{tracker.openbittorrent.com}, \\\url{tracker.publicbt.com}, and \\\url{https://s3-us-east-1-elb.amazonaws.com}.

The first URL contacted is router.bittorrent.com which acts as a registration point for the client and a way to initiate DHT participation. \texttt{update.browser.bittorrent.com} runs a script on connection which checks the version of the client connecting and generates a redirect to download the relevant update if applicable. The update check is triggered when the local client sends a \texttt{GET /windows/latest.json} request and, if an update is required the response contains a URL \url{http://update.browser.bittorrent.com/windows/<latest version number>/mini_installer.exe}. This secondary installer silently deploys a new folder alongside the original version with the new version number and the preferences and path files are updated accordingly.

As part of the installation procedure Project Maelstrom installs the following executables to the \texttt{\textbackslash Application\textbackslash}  directory:\\
maelstrom.exe (desktop shortcut target)\\ \texttt{MD5: 44d2641129cc922c5bc3545db8c8cda7}\\
chrome.native.torrent.exe(browser torrent manager)\\ \texttt{MD5: 85871a540cd77a3c419658fbe21c682b}\\
These executables are copies of files stored in the \texttt{<version>} subfolder and any update will cause the current files to be renamed with a prefix of ``old$\_$'' before being replaced with the later version. The file \texttt{VisualElementsManifest.xml} contains references to logos and other application images that include the <version> number to indicate which is the current active subfolder though unless manually altered this will usually indicate the latest version installed.

The subfolder \texttt{\textbackslash Application\textbackslash <version>\textbackslash webui} contains the base files for displaying default Chromium fonts, backgrounds, borders as well as all the files necessary for the start page and the \texttt{index.html} and images for the Project Maelstrom ``onboarding'' homepage partly depicted in figure\ref{fig:startpage}

In addition to the \texttt{Application} folder\\
\texttt{AppData\textbackslash Local\textbackslash Maelstrom\textbackslash User Data\textbackslash} is provisioned to store user specific data including user preferences, browser and torrent settings and browsing history. The majority of files are stored in the \texttt{Default} folder which contains a subset of the standard Chrome installation files in addition to files specific to Project Maelstrom. These will be covered in section \ref{ffiles} with the other files of forensic value to an investigator.

The second part of the installation involves the deployment of the \texttt{BitTorrent Maelstrom} folder in \texttt{\textbackslash Users \textbackslash <username> \textbackslash Appdata\textbackslash Roaming\textbackslash} to handle all torrent and DHT related activities. Any new torrents requested are saved here along with the corresponding .resume file. Torrent content is stored in the \texttt{cache} folder to speed up future access to to act as a repository to be shared with other users. On installation the root folder is used to store the files\\
\resizebox{\linewidth}{!}{%
\texttt{8E65684D700ECC41A09A60EE58991845EA56F734.resume}}\\
\resizebox{\linewidth}{!}{%
\texttt{8E65684D700ECC41A09A60EE58991845EA56F734.torrent}}\\
which are the torrent files associated with the Project Maelstrom Startpage.

\subsection{Settings}
\label{set}
Being built on Chromium, Project Maelstrom has all of the same browsing options available to the user. A standard installation has the same browser defaults set as a standard Chrome installation. One setting of interest is found in \texttt{Settings > Advanced Settings > System}. The option to ``continue running background apps when Maelstrom is closed'' is enabled by default. This will result in the \texttt{chrome.native.torrent.exe} executable running and sharing cached website torrent data from the users system without any visible indicator. In testing the data rate was observed to maintain at an average 800 bytes per second up and 800 bytes per second down. This was most likely just DHT updates and ping traffic. At any given time there were approximately 50 concurrent connections recorded.

The Project Maelstrom specific settings can be found in the ``Torrent'' section of the settings page. Each settings is customisable but at the time of installation the defaults are:
\begin{itemize}[noitemsep,topsep=0pt]
\item{Cache Size} -- The amount of space available to store cached torrent data is set to 5GB with a warning not to exceed 100GB.
\item{Sharing Ratio} -- A slider can be used to change the ratio from 0 upload to unlimited upload. The default setting is upload data is equal to the amount downloaded. This sharing ratio is measured on a per torrent basis and is presented to the user as a measure of contribution.
\item{Rate Limits} -- The default setting is to not restrict upload or download speeds.
\item{Transfer Limit} -- The default setting is not to impose a cap on total transfer amounts.
\item{Port Settings} -- A port is selected at installation as the static communication port. This port number can be customised here or can be set to be randomly selected.
\item{Proxy Server} -- The proxy settings for torrent transfers can be selected from a dropdown list including SOCKS which will allow for the use of TOR for added anonymity. By default, no proxy is selected.
\item{Privacy} -- By default the option to send anonymous crash data to BitTorrent Inc. is enabled.
\end{itemize}

\subsection{Files of Forensic value}
\label{ffiles}
During the first run the settings and preferences files stored in \texttt{User Data} and  \texttt{Application} are populated. at the same time, the torrent specific settings and discovery are recorded in the Roaming directory. Because of Project Maelstrom's roots in the Chromium browser, many of the same forensically important resources are present. However, the dual nature of the utility does present some degree of difference.For example, while the History tab shows the \texttt{bittorrent://} and \texttt{magnet://} sites visited, the content for these sites is not stored in the standard caches or data files. Instead checking the output of \texttt{about:cache} in the browser will only show standard website files. All caching and management relating to torrents are stored in the \texttt{Roaming\textbackslash BitTorrent Maelstrom} folder.
\begin{itemize}[noitemsep,topsep=0pt]

\item{\texttt{Local\textbackslash Maelstrom\textbackslash Application}} -- This directory contains initial setup files and content added during installation. Of most use is confirmation of an active installation, timestamps to indicate install time and a record of the initial settings.
\item{\texttt{Local\textbackslash Maelstrom\textbackslash User Data}} -- contains the majority of the user specific settings and activity records. This folder and its subfolders can be examined using established forensic techniques for use on Chrome browsers but also includes some new files that may be of use.
\begin{enumerate}[noitemsep,topsep=0pt]

\item{The history SQLite3 database file stored in the \texttt{default} folder will contain both the \texttt{magnet:?} URI and the resolved \texttt{bittorrent://} address. The \texttt{magnet:?} entry will have a title (same as the torrent file), while the \texttt{bittorrent://} address will have the title of the index.html page contained in the bundle.}
\item{Origin Bound Certs will only contain the certificates specific to the normal internet browsing activities.}
\item{The HTML rendered after scripts have been run is stored in the \texttt{Cache} directory. for example, the full HTML of \texttt{bittorrent://welcome} is stored as part of \texttt{f$\_$000005}.}
\item{\texttt{Local Storage} -- contains persistent information used by visited websites. From the first run this will contain the Unique ID used by bench.utorrent.com to gather anonymous usage statistics. The contents of this folder are SQLite3 databases with individual data sets stored as blobs}
\item{\texttt{Session Storage} -- contains a list of BitTorrent UUIDs. The first UUID matches that stored in the \texttt{bench.utorrent} local storage file. The remainder are those of remote peers the browser is in contact with.}
\end{enumerate}
\item{\texttt{Roaming\textbackslash BitTorrent Maelstrom\textbackslash}} Contains the bulk of the torrent handling and storage elements of Project Maelstrom. The root folder has many items of forensic value including:
\begin{enumerate}[noitemsep,topsep=0pt]

\item{\texttt{<bt.infohash>.torrent}} -- The torrent file used to locate and download resources from peers. This is a standard torrent file format listing all sub-files contained within the ``bundle'' and the hashes for each piece and the trackers.
\item{\texttt{<bt.infohash>.resume}} -- A resume file created to allow .torrent data processing to be paused and continued without having to restart to ensure full data transfer. The creation of the resume file could be used as a guideline to when the related .torrent was first processed.
\item{\texttt{dht.dat}} -- This consists of a bencoded list of observed IP:Port combinations of peers participating in the DHT. The content starts with the value \texttt{id20:}, which is followed by the BitTorrent client's unique ID on the DHT network. It is then followed by the number of observed nodes or peers and a listing of the IP:Port pairs in 6 byte IPv4 representations.
\item{\texttt{settings.dat}} -- This file contains the torrent client settings and the usage statistics.
\item{\texttt{Cache}} -- This stores the downloaded torrent files, which correspond to the name of the torrent appended with the \texttt{btinfohash} of the piece. Scripts and web pages are loaded and executed from this directory to improve browsing speed for the user and to provide a repository to share to other peers.
\item{\texttt{trusted} folder} -- This stores the \texttt{bittorrent.crt} file for use in peer secure transfer negotiation.
\end{enumerate}
\end{itemize}

\subsection{Local File Remnants}
\label{filerem}

While there is no dedicated uninstall executable for Project Maelstrom, the registry key\\ \texttt{HKCU\textbackslash <user>\textbackslash Software\textbackslash Microsoft\textbackslash Windows\textbackslash \\ CurrentVersion\textbackslash Uninstall\textbackslash Maelstrom UninstallString}\\ shows the control panel remove program options to be the equivalent of running\\ \texttt{``AppData\textbackslash Local\textbackslash Maelstrom\textbackslash Application\textbackslash\\ <version> \textbackslash Installer\textbackslash setup.exe'' $--$uninstall}\\
By default the option to remove user history is not checked. If left as default, all of the files and folders in \texttt{AppData\textbackslash Local\textbackslash Maelstrom\textbackslash User Data\textbackslash} is left intact including any history and cookie folders as well as user specific settings and preferences.

If the option to remove User history is selected then the entire \texttt{AppData\textbackslash Local\textbackslash Maelstrom\textbackslash } folder and all subfolders are removed. However, in either scenario the \texttt{AppData\textbackslash Roaming\textbackslash BitTorrent Maelstrom\textbackslash } folder is left untouched along with all of the subfolders and files contained therein including \texttt{dht.dat} and any \texttt{.torrent} and \texttt{.resume} files that may have been stored. Additionally all cached data held in the \texttt{cache} directory is left intact and retrievable.

\section{Project Maelstrom: Protocol Analysis}
\label{protocol}

\subsection{Accessing a Website}
Project Maelstrom supports all of the regular protocols that ship with Chromium, alongside two BitTorrent ecosystem specific URIs; \texttt{bittorrent://} and \texttt{magnet:?}.

In order to load any given webpage, the browser must:\\
\begin{enumerate}[nolistsep,topsep=0pt]
\item Resolve the \texttt{magnet} link to a \texttt{.torrent} file.
\item Identify other peers currently serving that webpage.
\item Identify any torrents contained within the webpage bundle.
\item Cache part files and store them as blocks of the whole in sequence.
\item Process any scripts that determine content.
\item Update the sharing status of the local content to include the new webpage.
\end{enumerate}

\subsubsection{Peer Discovery}

In order to discover an initial set of peers the client contacts \texttt{router.bittorrent.com}. The packet requesting peers contains the command ``find\_nodes'' and is bencoded with the Local Peer ID followed by the BitTorrent share ID. In response a bencoded list of registered peers is returned, each with a 26 byte entry consisting of a 20 byte PeerID and a 6 byte IP:Port entry. The local client then uses DHT$\_$Ping to discover live hosts and standard torrent discovery is performed as described in the BitTorrent Extension Protocols (BEP) -- specifically BEP005 deals with DHT discovery and bootstrapping \citep{loewenstern2008bittorrent}. Any peer that has the \texttt{chrome.native.torrent.exe} executable running, even if the Maelstrom browser is shutdown, that has the torrent in its cache will become available as valid content source.

\subsubsection{Content Negotiation}
Links in Project Maelstrom are handled either as torrent files or as \texttt{magnet:?} links. If a magnet link is followed the Maelstrom browser must first resolve the data provided in the link to a valid torrent file. This is achieved either by looking up the tracker URL included in the magnet link using the \texttt{tr} parameter or one of the default trackers defined at installation. Once discovered the torrent ID is then used to construct \texttt{find$\_$nodes} requests that are sent to the known DHT peers and \texttt{router.bittorrent.com}.

\subsection{Maelstrom Development}
The development of websites for distribution through Maelstrom does not require any special consideration during the development of the website content, i.e., the HTML, CSS, JavaScript, graphics, multimedia content, etc. Where the traditional web development process differs is with regards to the uploading of content to a web server. These websites can be delivered either as a single large torrent resource broken into pieces or, for ease of maintenance and to avoid users having to re-download the entire site every time the content is changed, as a series of .torrent files linked by an encompassing ``Bundle''. 

To make the process more streamlined, BitTorrent Inc. have developed a number of helper tools and guidelines \citep{torrentwebtools}.

\section{Conclusion}
\label{conclusion}

It is envisioned that the popularity of Project Maelstrom (and similar decentralised alternatives) will increase in the coming months and years due to its ease of use, easy web-authoring and zero-cost model (to content creators and consumers alike). Perhaps BitTorrent Inc. will have the ability to raise their new Internet from the deep web into the hands of regular Internet users.

\subsection{Future Work}
\label{future}
With the browser just having been recently released as a public beta, there remains much to be discovered regarding the nuances of this decentralised web. Some areas for future work include:

\begin{itemize}[noitemsep,topsep=0pt]

\item Development of a tool capable of monitoring who is accessing any specific Maelstrom-only website. The decentralised nature of the protocol leaves it vulnerable for third-parties to garner visitor information statistics that are normally reserved for the administrator of the web host such as location, access duration, files downloaded, frequency of repeat visits, content available for sharing, etc.
\item Perform an analysis of ZeroNet \citep{zeronet}, an open source Project Maelstrom alternative. ZeroNet does not require the use of a specific browser -- instead running as a service on the local machine.
\end{itemize}




{\footnotesize \bibliographystyle{plainnat}
\bibliography{./bibliography}}

\begin{thebibliography}{12}
\providecommand{\natexlab}[1]{#1}
\providecommand{\url}[1]{\texttt{#1}}
\expandafter\ifx\csname urlstyle\endcsname\relax
  \providecommand{\doi}[1]{doi: #1}\else
  \providecommand{\doi}{doi: \begingroup \urlstyle{rm}\Url}\fi

\bibitem[Biryukov et~al.(2013)Biryukov, Pustogarov, and Weinmann]{Biryukov2013}
Alex Biryukov, Ivan Pustogarov, and Ralf-Philipp Weinmann.
\newblock Trawling for tor hidden services: Detection, measurement,
  deanonymization.
\newblock In \emph{Proceedings of the 2013 IEEE Symposium on Security and
  Privacy}, pages 80--94, 2013.

\bibitem[{BitTorrent Inc.}(2015)]{torrentwebtools}
{BitTorrent Inc.}
\newblock {Torrent Web Tools}.
\newblock \url{https://github.com/bittorrent/torrent-web-tools}, May 2015.

\bibitem[Choi and Perez(2007)]{choi2007online}
David~Y Choi and Arturo Perez.
\newblock Online piracy, innovation, and legitimate business models.
\newblock \emph{Technovation}, 27\penalty0 (4):\penalty0 168--178, 2007.

\bibitem[Klinker(2014)]{klinker2014project}
Eric Klinker.
\newblock {Project Maelstrom: The Internet We Build Next}.
\newblock
  \url{http://blog.bittorrent.com/2014/12/10/project-maelstrom-the-internet-we-build-next/},
  December 2014.

\bibitem[Klinker(2015)]{klinker2015project}
Eric Klinker.
\newblock {Project Maelstrom Enters Beta}.
\newblock
  \url{http://blog.bittorrent.com/2015/04/10/project-maelstrom-enters-beta/},
  April 2015.

\bibitem[Loewenstern and Norberg(2008)]{loewenstern2008bittorrent}
Andrew Loewenstern and Arvid Norberg.
\newblock {DHT Protocol}.
\newblock \\\url{http://www.bittorrent.org/beps/bep_0005.html}, 2008.
\newblock [Online; accessed July 2015].

\bibitem[Roos et~al.(2014)Roos, Schiller, Hacker, and
  Strufe]{roos2014measuring}
Stefanie Roos, Benjamin Schiller, Stefan Hacker, and Thorsten Strufe.
\newblock Measuring freenet in the wild: Censorship-resilience under
  observation.
\newblock In \emph{Privacy Enhancing Technologies}, pages 263--282. Springer,
  2014.

\bibitem[Scanlon et~al.(2014)Scanlon, Farina, Le~Khac, and
  Kechadi]{scanlon2014leveraging}
Mark Scanlon, Jason Farina, Nhien-An Le~Khac, and M-Tahar Kechadi.
\newblock {Leveraging Decentralisation to Extend the Digital Evidence
  Acquisition Window: Case Study on BitTorrent Sync}.
\newblock \emph{{Journal of Digital Forensics, Security and Law}}, pages
  85--99, 2014.

\bibitem[Scanlon et~al.(2015)Scanlon, Farina, and Kechadi]{Scanlon2015}
Mark Scanlon, Jason Farina, and M-Tahar Kechadi.
\newblock {Network Investigation Methodology for BitTorrent Sync: A
  Peer-to-Peer Based File Synchronisation Service}.
\newblock \emph{Computers \& Security}, 2015.
\newblock \doi{http://dx.doi.org/10.1016/j.cose.2015.05.003}.

\bibitem[Watters et~al.(2011)Watters, Layton, and Dazeley]{Watters201179}
Paul~A. Watters, Robert Layton, and Richard Dazeley.
\newblock {How much material on BitTorrent is infringing content? A case
  study}.
\newblock \emph{Information Security Technical Report}, 16\penalty0
  (2):\penalty0 79 -- 87, 2011.

\bibitem[Zantout and Haraty(2011)]{zantout2011i2p}
Bassam Zantout and Ramzi Haraty.
\newblock I2p data communication system.
\newblock In \emph{ICN 2011, The Tenth International Conference on Networks},
  pages 401--409, 2011.

\bibitem[{ZeroNet}(2015)]{zeronet}
{ZeroNet}.
\newblock \url{https://github.com/HelloZeroNet/ZeroNet}, May 2015.

\end{thebibliography}


\end{document}